\documentclass[final,5p,times,twocolumn,sort&compress]{elsarticle}

\usepackage{amssymb}
\usepackage{bm}
\usepackage{amssymb,amsbsy,amsmath,amsfonts}
\usepackage{epsfig}
\usepackage{graphicx}

\journal{Physics Letters B}
\newcommand{\be}{\begin{equation}}
\newcommand{\ee}{\end{equation}}
\newcommand{\bea}{\begin{eqnarray}}
\newcommand{\eea}{\end{eqnarray}}
\newcommand{\bef}{\begin{figure}}
\newcommand{\eef}{\end{figure}}
\newcommand{\bce}{\begin{center}}
\newcommand{\ece}{\end{center}}

\def\mev{\rm{ MeV}}
\def\gev{\rm{ GeV}}
\def\fm{\rm{ fm}}

\newcommand{\bD}{{\bar D}}
\newcommand{\ignore}[1]{}

\begin{document}

\begin{frontmatter}



\title{Formation spectra of charmed meson--nucleus systems using an
  antiproton beam}


\author{J. Yamagata-Sekihara}
\address{National Institute of Technology, Oshima College, Oshima,
  Yamaguchi, 742-2193, Japan}

\author{C. Garcia-Recio}
\address{Departamento de F{\'\i}sica At\'omica, Molecular y
  Nuclear, and Instituto Carlos I de F{\'i}sica Te\'orica y
  Computacional, Universidad de Granada, E-18071 Granada, Spain}

\author{J. Nieves}
\address{Instituto de F{\'\i}sica Corpuscular (centro mixto
  CSIC-UV), Institutos de Investigaci\'on de Paterna, Aptdo.\ 22085,
  46071, Valencia, Spain}

\author{L.L. Salcedo}
\address{Departamento de F{\'\i}sica At\'omica, Molecular y
  Nuclear, and Instituto Carlos I de F{\'i}sica Te\'orica y
  Computacional, Universidad de Granada, E-18071 Granada, Spain}

\author{L. Tolos}
\address{Instituto de Ciencias del Espacio (IEEC/CSIC), Campus
  UAB, Carrer de Can Magrans s/n, 08193 Cerdanyola del Valles, Spain}

\address{Frankfurt Institute for Advanced Studies, Johann Wolfgang
  Goethe University, Ruth-Moufang-Str.\ 1, 60438 Frankfurt am Main,
  Germany}

\begin{abstract}
  We investigate the structure and formation of charmed meson--nucleus
  systems, with the aim of understanding the charmed meson--nucleon
  interactions and the properties of the charmed mesons in the nuclear
  medium.  The $\bar{D}$ mesic nuclei are of special interest, since
  they have tiny decay widths due to the absence of strong decays for
  the $\bar{D} N$ pair.  Employing an effective model for the $\bar{D} N$ and
  $D N$ interactions and solving the Klein--Gordon equation for
  $\bar{D}$ and $D$ in finite nuclei, we find that the
  $D^{-}$-${}^{11}\rm{B}$ system has $1 s$ and $2 p$ mesic nuclear
  states and that the $D^{0}$-${}^{11}\rm{B}$ system binds in a $1
  s$ state.  In view of the forthcoming experiments by the PANDA and
  CBM Collaborations at the future FAIR facility and the J-PARC
  upgrade, we calculate the formation spectra of the $[D^{-}$-${}^{11}\rm{B}]$ and $[D^{0}$-${}^{11}\rm{B}]$ mesic nuclei for an
  antiproton beam on a ${}^{12} \rm{C}$ target. Our results suggest
  that it is possible to observe the $2 p$ $D^{-}$ mesic nuclear state
  with an appropriate experimental setup.

\end{abstract}

\begin{keyword}
charmed mesic nuclei, formation spectra, $DN$ and $\bar DN$ interaction, Klein-Gordon equation, Green's function method

\PACS{21.85.+d, 14.40.Lb, 21.65.Jk, 36.10.Gv}


\end{keyword}

\end{frontmatter}


\section{Introduction}
\label{sec:intro}

The study of hadronic atoms provide essential information on the
properties of hadron-nucleon interactions, hadrons in matter as well
as the properties of nuclei that are not accessible by other
probes. Pionic and kaonic atoms have been intensively investigated
over the years \cite{Ericson:1966fm,Friedman:2007zza,
  Nieves:1993ev,GarciaRecio:1991wk,Hirenzaki:2008zz,Gilg:1999qa},
whereas antiprotons in atoms have become a matter of recent interest
\cite{Wycech:2007jb,Klos:2007is,Trzcinska:2001sy}.

In view of the forthcoming experiments by the PANDA and CBM
Collaborations at the future FAIR facility \cite{fair} and the J-PARC
upgrade \cite{jparc}, the attention has been also focused on charmed
meson-nucleus systems.  One of the first works on charmed mesic nuclei
analyzed the possibility of $D^-$ atoms \cite{Tsushima:1998ru}. There,
the $1s$, $2s$ and $1p$ states of $D^-$ in $^{208}$Pb were evaluated
using the quark-meson coupling model of
Ref.~\cite{Guichon:1987jp}. The energy levels of the $\bD$ meson in
$^{208}$Pb and $^{40}$Ca were obtained in \cite{Yasui:2012rw} within a
model for the charmed meson-nucleon interaction based on the pion
exchange. Also, $\bD NN$ and $\bD^* NN$ bound states were predicted in
\cite{Yasui:2009bz,Yamaguchi:2013hsa} as well as a bound state of
$DNN$ in \cite{Bayar:2012dd}.

All these works rely upon building a realistic charmed meson-nucleon
interaction and extending the analysis to the nucleus. In that
respect, unitarized meson-baryon coupled-channel approaches including
the charm degree of freedom have been very successful
\cite{Tolos:2004yg, Tolos:2005ft, Lutz:2003jw, Lutz:2005ip,
  Hofmann:2005sw, Hofmann:2006qx, Lutz:2005vx, Mizutani:2006vq,
  Tolos:2007vh, JimenezTejero:2009vq,
  JimenezTejero:2011fc,Haidenbauer:2007jq, Haidenbauer:2008ff,
  Haidenbauer:2010ch, Wu:2010jy, Wu:2010vk, Wu:2012md,
  Oset:2012ap}. However, these models do not explicitly incorporate
heavy-quark spin symmetry (HQSS)\cite{IW89,Ne94,MW00} and, thus, it is
unclear whether they fulfilled the constraints imposed by HQSS. HQSS
is a QCD symmetry that appears when the quark masses, such as the
charm mass, become larger than the typical confinement scale.

The implementation of HQSS constraints on the meson-baryon
interactions with heavy-quark degrees of freedom has been more
recently studied in
\cite{GarciaRecio:2008dp,Gamermann:2010zz,Romanets:2012hm,GarciaRecio:2012db,Garcia-Recio:2013gaa,Tolos:2013gta,Xiao:2013yca,Ozpineci:2013zas}. Among
these works, we must highlight those based on an extension of the
Weinberg-Tomozawa (WT) interaction to spin-flavor including HQSS
constraints
\cite{GarciaRecio:2008dp,Gamermann:2010zz,Romanets:2012hm,GarciaRecio:2012db,Garcia-Recio:2013gaa,Tolos:2013gta}.
Within this approach, we have analyzed the properties of $D$ and $\bD$
as well as $D^*$ and $\bD^*$ in dense matter and studied the formation
of charmed-meson nucleus bound states
\cite{Tolos:2009nn,GarciaRecio:2010vt,GarciaRecio:2011xt}.

In Ref.~\cite{GarciaRecio:2010vt} we have obtained that $D^0$ binds
weakly with nuclei, in contrast to \cite{Tsushima:1998ru}, while the
$D^0$-nucleus states have significant widths, in particular for heavy
nuclei such as $^{208}$Pb. The best chances for observation of bound
states are in the region of $^{24}\mbox{Mg}$, provided an orbital
angular momentum separation can be done. Moreover, only $D^0$-nucleus
bound states are possible since the Coulomb interaction prevents the
formation of observable bound states for $D^+$ mesons. With regards to
$\bar D$-mesic nuclei, not only $D^-$ but also $\bar{D}^0$ bind in
nuclei \cite{GarciaRecio:2011xt}. The spectrum contains states of
atomic and of nuclear types for all nuclei for $D^-$ whereas only nuclear states are present for $\bar{D}^0$ in
nuclei, as
expected. Compared to the pure Coulomb levels, the atomic states are
less bound. The nuclear ones are more bound and may present a sizable
width. Moreover, nuclear states only exist for low angular momenta.

In this work, we continue these previous studies and investigate the
possibility of observing $D^{-}$-${}^{11}\rm{B}$ and
$D^{0}$-${}^{11}\rm{B}$ bound states in ${}^{12} \rm{C} (
\bar{p}, D^{+})$ and ${}^{12} \rm{C} ( \bar{p}, \bar{D}^{0})$
nuclear reactions.  The formation spectra are calculated with the Green's
function method.  This is the first attempt to calculate the formation
spectra for charmed mesic nuclear states with an energy dependent
optical potential coming from the charmed meson-nucleon interaction in
matter.  Throughout this study, we set the incident antiproton beam at
$8 \gev /c$ and the final-state $D^{+}$ and $\bar{D}^{0}$ mesons to go
forward direction, in an attempt to give useful information to
experiments with antiprotons beams, such as PANDA (FAIR) and J-PARC.

This paper is organized as follows.  In Sec.~\ref{sec:form} we present
the formation spectra in terms of the production cross sections of the
$( \bar{p}, D^{+})$ and $( \bar{p}, \bar{D}^{0})$ reactions on a
nuclear target.  We also describe the $D N$ and $\bar{D} N$ effective
interactions in medium together with the $D$ and $\bar D$
self-energies and optical potentials used in this study. Details on the construction of the nuclear density to evaluate  the optical potentials for a given nucleus are presented in \ref{appendix}.  Next, in
Sec.~\ref{sec:results} we show our numerical results on the structure
of charmed meson-nuclear systems and the formation spectra for them.
Section~\ref{sec:conc} is devoted to the conclusions of this paper.

\section{Formalism}
\label{sec:form}

First of all, we present the formalism for the formation spectra of
charmed meson-nuclear systems in terms of the differential cross
sections of the $( \bar{p}, D^{+})$ and $( \bar{p}, \bar{D}^{0})$
reactions on a nuclear target.  In the calculation of the formation
spectra, the charmed meson--nucleon scattering amplitudes as well as
the charmed meson self-energies and optical potentials are essential.
In this study we employ the approach proposed in
Refs.~\cite{Tolos:2009nn,GarciaRecio:2010vt,GarciaRecio:2011xt}. The
charmed meson-nucleon interaction in matter and the corresponding
charmed meson self--energies are presented in Sec.~\ref{sec:amp}.
Then, in Sec.~\ref{sec:spec} we construct the optical potential for
the charmed mesons, that are needed for the solution of the
 Klein--Gordon equation (KGE) for the charmed meson-nuclear systems, and
summarize our procedure to calculate the production cross section in
the Green's function method.

\subsection{Charmed meson--nucleon scattering amplitudes and charmed meson self-energies}
\label{sec:amp}

The different charmed meson--nucleon scattering amplitudes in
symmetric nuclear matter and the corresponding charmed meson
self-energies are obtained following a self-consistent procedure in
coupled channels, as described in
\cite{Tolos:2009nn,GarciaRecio:2010vt} for the $D$ meson and in
\cite{GarciaRecio:2011xt} for $\bar D$ meson. Here we summarize the
main features.

The $s$-wave transition charmed meson--nucleon potential of the
Bethe-Salpeter equation is derived from an effective Lagrangian that
implements HQSS \cite{IW89,Ne94,MW00}.  HQSS is an approximate QCD
symmetry that treats on equal footing heavy pseudoscalar and vector
mesons, such as charmed and bottomed mesons
\cite{GarciaRecio:2008dp,Gamermann:2010zz,Romanets:2012hm,GarciaRecio:2012db,Garcia-Recio:2013gaa,Tolos:2013gta,Tolos:2009nn,GarciaRecio:2010vt,GarciaRecio:2011xt,Xiao:2013yca,Ozpineci:2013zas}. The
effective Lagrangian includes the lowest-lying pseudoscalar and vector
mesons as well as $1/2^+$ and $3/2^+$ baryons. It reduces to the WT
interaction term in the sector where Goldstone bosons are involved and
incorporates HQSS in the sector where heavy quarks participate. This
SU(6)$\times$HQSS model is justified in view of the reasonable
semi-qualitative outcome of the SU(6) extension
\cite{Gamermann:2011mq} and on a formal plausibleness on how the SU(4)
WT interaction in the heavy pseudoscalar meson-baryon sectors comes
out in the vector-meson exchange picture (see for instance Refs.~\cite{Lutz:2005ip, Mizutani:2006vq}).

The extended WT meson-baryon interaction in the coupled meson-baryon
basis with total charm $C$, strangeness $S$,
isospin $I$ and spin $J$, is given by
\begin{eqnarray}
&&V_{ij}^{CSIJ}(\sqrt{s}) =\nonumber \\
&&D_{ij}^{CSIJ}\,\frac{2\sqrt{s}-M_i-M_j}{4f_if_j} 
\sqrt{\frac{E_i+M_i}{2M_i}}\sqrt{\frac{E_j+M_j}{2M_j}}, 
\label{eq:pot}
\end{eqnarray}
where $\sqrt{s}$ is the center of mass (C.M.) energy of the system; $E_i$ and
$M_i$ are, respectively, the C.M. on-shell energy and mass of the baryon in the channel
$i$; and $f_i$ is the decay constant of the meson in the $i$-channel.
Symmetry breaking effects are introduced by using physical masses and decay
constants.  The $D_{ij}^{CSIJ}$ are the matrix elements coming from the group structure of
the extended WT interaction.  

The amplitudes in nuclear matter, $T^{\rho,CSIJ}(P^0,{\bm P})$ with $P=(P^0,{\bm P})$ the total four-momentum, are obtained by
solving the on-shell Bethe-Salpeter equation with the tree level amplitude
$V^{CSIJ}(\sqrt{s})$:
\begin{eqnarray}
&&T^{\rho,CSIJ}(P) = \nonumber \\
&& \frac{1}{1-V^{CSIJ}(\sqrt{s})\, G^{\rho,CSIJ}(P)}\,V^{CSIJ}(\sqrt{s})
,
 \label{eq:scat-rho}
\end{eqnarray}
where the diagonal $G^{\rho,CSIJ}(P)$ matrix accounts for the
charmed meson--baryon loop in nuclear matter \cite{Tolos:2009nn,GarciaRecio:2011xt}.
We focus in the non-strange $S=0$ and singly charmed
$C=1$ sector, where $DN$ and $D^*N$ are embedded, as well as the $C=-1$ one, with $\bD N$ and $\bD^* N$. \footnote{Note that $D$ denotes $D^+$ and $D^0$, whereas $\bD$ indicates $\bD^-$ and $\bD^0$}

The $D(\bD)$ and $D^*(\bD^*)$ self-energies in symmetric nuclear matter, $\Pi(q^0,{\bm q}; \rho)$, are  obtained by summing
the different isospin transition amplitudes for $D(\bD)N$ and $D^*(\bD^*)N$ over the nucleon Fermi
distribution, $p_F$. For the $D(\bD)$  we have
\begin{eqnarray}
&&\Pi_{D(\bD)}(q^0,{\bm q}; \rho) = \nonumber \\ 
&&\int_{p \leq p_F} 
\frac{d^3p}{(2\pi)^3} \,
   \Big[\, T^{\rho,0,1/2}_{D(\bD) N} (P^0,{\bm P}) +
3 \, T^{\rho,1,1/2}_{D(\bD) N}(P^0,{\bm P}) \Big]
, \ \ \ \
\label{eq:selfd}
\end{eqnarray}
while for $D^*(\bD^*)$
\begin{eqnarray}
&&\Pi_{D^*(\bD^*)}(q^0,{\bm q}; \rho\,) = \nonumber \\  
&&\int _{p \leq p_F} \frac{d^3p}{(2\pi)^3} \,
\Bigg[ \frac{1}{3} \, T^{\rho,0,1/2}_{D^*(\bD^*) N}(P^0,{\bm P}) +
T^{\rho,1,1/2}_{D^*(\bD^*) N}(P^0,{\bm P}) 
\nonumber \\
&& 
+   \frac{2}{3} \,
T^{\rho,0,3/2}_{D^*(\bD^*) N}(P^0,{\bm P}) + 
2 \, T^{\rho,1,3/2}_{D^*(\bD^*) N}(P^0,{\bm P})\Bigg]
. 
\label{eq:selfds}
\end{eqnarray}
\noindent
In the above equations, $P^0=q^0+E_N({\bm p})$ and ${\bm
P}={\bm q}+{\bm p}$ are the total energy and momentum of the
meson-nucleon pair in the nuclear matter rest frame, and $(q^0,{\bm
q})$ and $(E_N,{\bm p})$ stand for the energy and momentum of the
meson and nucleon, respectively, in that frame. Those self-energies are determined self-consistently since they
are obtained from the in-medium amplitudes which contain the meson-baryon loop
functions, and those quantities themselves are functions of the self-energies.

\subsection{Optical potential and formation spectra}
\label{sec:spec}

In order to calculate the formation spectra of the meson--nucleus
bound states,  we need the optical potential of a meson in the nucleus. Relying on the local density approximation, 
 we evaluate the optical
potential for the  $D^{-}$ ($D^{0}$) mesons \footnote{For the calculation of the $D^{-}$ optical potential we do not vary the
subtraction point, namely $\alpha=1$, and we do not consider the nucleon extraction energy (or gap) (see \cite{GarciaRecio:2011xt} for details).}
\begin{equation}
  V_{\rm opt} ( r, E_D) = \frac{1}{2m_D}
  \Pi _{D^{-} (D^{0})} (E_D,{\bm q}=0; \rho ( r ) ) ,
\end{equation}
where $r$ is the distance from the center of the nucleus and $E_D$ is the
energy of the charmed meson, i.e. $E_D$ is same as $q^0$ in Sec. IIA. 
The nuclear density 
$\rho (r)$ is evaluated  from the
neutron and proton densities. The densities are deconvoluted as in Ref.~\cite{Nieves:1993ev} to
account for the proton and neutron finite sizes.  The details on the proton and neutron densities are given in \ref{appendix}.

With the optical potential $V_{\rm opt}$ we can obtain the meson wave
function in the nucleus by solving the KGE
\begin{eqnarray}
  \label{KG_eq}
        &&[ - \nabla ^{2} + \mu ^{2} + 2 \mu V_{\rm opt} (r , E_D- V_{\rm coul}(r) ) ]
        \phi( {\bm r} ) \nonumber \\
        &&= [ E_D - V_{\rm coul}(r)]^2 \phi( {\bm r} ) .
\end{eqnarray}
with $\mu$ the $D$ meson-nucleus reduced mass. Here $V_{\rm coul}(r)$
is the Coulomb potential given by
\begin{equation}
  \label{V_coul}
  V_{\rm coul} (r) = e^{2} Z_{D} \int
  \frac{\rho_{\rm ch}( r^{\prime} )}
       {\left | {\bm r}-{\bm r}^{\prime} \right |}
       d^3 r^{\prime} , 
\end{equation}
\noindent
where $e$ is the elementary charge, $Z_{D}$ is the charge of the
charmed meson, and $\rho_{\rm ch} ( r )$ is the charge distribution of
the nucleus of  Eq.~\eqref{rho_ch}.  We note that, for the
$D^{0}$-nucleus system, the Coulomb interaction is automatically
removed, since $Z_{D} = 0$.

Next we discuss the procedure to calculate the formation spectra in terms of the
differential cross sections in the Green's function
method~\cite{Morimatsu:1985pf}.  The details of the
Green's function method can be found in Refs.~\cite{Hayano:1998sy,
  Klingl:1998zj, Jido:2002yb, Nagahiro:2003iv, Nagahiro:2005gf}, and
here we only summarize the main features.  

In this work, we calculate
the formation spectra of the $D^{-}$-, $D^{0}$-${}^{11}\rm{B}$ systems in the 
\begin{eqnarray}
\bar{p} + {}^{12}\rm{C} &\to& \left[ {}^{11}\rm{B}-D^-
  \right]+D^+ \label{eq:rea1}\\
\bar{p} + {}^{12}\rm{C} &\to& \left[ {}^{11}\rm{B}-D^0  \right]+\bar{D}^0
\label{eq:rea2}
\end{eqnarray}
reactions. As mentioned in the introduction, these processes are of interest for the forthcoming
experiments by the PANDA and CBM Collaborations at the future FAIR
facility and in J-PARC. For simplicity, we concentrate on the
formation spectra of the $(\bar{p}, \bar{D})$ process.

The present method starts with the separation of the  cross
section into the nuclear response function $S(E_D)$ and the elementary
cross section for the $p({\bar p}, {\bar D})D$ reaction within the impulse
approximation for $D$ meson production
\begin{equation}
\label{crossS}
\left (
\frac{d^{2} \sigma}{d \Omega d E_D} \right )
_{ A ( {\bar p}, {\bar D} ) ( A - 1 ) \otimes D }
=\left ( \frac{d \sigma}{d \Omega} \right )^{\rm LAB}
_{p ( {\bar p}, {\bar D} ) D} \times S(E_D).
\end{equation}
\noindent
The differential cross section of the elementary process $p({\bar
  p},{\bar D})D$ in the laboratory frame (LAB), $( d \sigma / d\Omega )^{\rm
  LAB}_{p({\bar p},{\bar D})D}$, can be evaluated using some appropriate
models or be taken from experimental data.  For this cross section, we
will use the theoretical results of Ref.~\cite{Haidenbauer:2014rva}.
The nuclear response function $S (E_D )$ contains
information on the dynamics between $D-$meson and the final $(A - 1)$
nucleus.  
To calculate the nuclear response function, we
employ the Green's function method.  Namely, the nuclear response
function with a complex potential is formulated in
Ref.~\cite{Morimatsu:1985pf} as
\begin{equation}
\label{S(E_D)}
S ( E_D ) = - \frac{1}{\pi} {\rm Im}\sum_{f}
\int d^{3} r d^{3} r^{\prime} \; \tau ^{\dagger}_{f} ( \bm{r} ) 
G(E_D; \bm{r}, \bm{r}^{\prime}) \tau _{f} ( \bm{r}^{\prime} ) ,
\end{equation}
where $\tau _{f}$ denotes the transition amplitude of the initial state
$\bar{p}+{}^{A}Z$ to the proton--hole final nucleus and the outgoing
$D-{\bar D}$ 
meson pair, and $G(E_D; \bm{r}, \bm{r}^{\prime})$ is the Green's function of
the $D$ meson interacting with the nucleus.  The summation is taken
over all possible final states $f$.  The Green's function $G(E_D;
\bm{r}, \bm{r}^{\prime})$ is defined as,
\begin{equation}
\label{Gfunc}
G ( E_D ; \bm{r}, \bm{r}^{\prime} )
= \langle \alpha | \phi( \bm{r} )
\displaystyle { \frac{1}{E_D - H_{D} + i\epsilon }}
\phi ^{\dagger} ( \bm{r}^{\prime} ) | \alpha \rangle ,
\end{equation}
where $\alpha$ indicates the proton--hole state, $H_{D}$ is the
Hamiltonian of the $D$ meson--nucleus system, and $\phi ^{\dagger} (
\bm{r} )$ is the $D$ meson creation operator.  The transition
amplitude $\tau _{f}$ involves the proton--hole wave function $\psi
_{j_{N}}$ and the distorted waves $\chi _{i}$ and $\chi _{f}$ of the
projectile and ejectile, respectively.  The distorted waves are
calculated within the eikonal approximation as
\begin{equation}
  \chi _{f}^{\ast} ( \bm{r} ) \chi _{i} ( \bm{r} )
  = \exp ( i \bm{q} \cdot \bm{r} ) F ( \bm{r} ) , 
\end{equation}
with the momentum transfer $\bm{q}$ and the distortion factor $F (
\bm{r} )$ defined as
\begin{equation}
  F ( \bm{r} ) = \exp \left [ - \frac{1}{2}
  \bar{\sigma} \int _{- \infty}^{\infty} d z^{\prime} \bar{\rho}
  ( z^{\prime} , \bm{r} ) \right ] , 
\end{equation}
where $\bar{\sigma}$ is the averaged  cross section
\begin{equation}
  \bar{\sigma} = \frac{\sigma _{\bar{p} N} + \sigma _{\bar D N}}{2} ,
\end{equation}
with  $\sigma _{\bar{p} N}$  
and $\sigma _{\bar D N}$,  the total $\bar{p} N$ and $D N$ cross
sections, respectively.  We use the values $\sigma
  _{\bar{p} N} = 59$ mb, and $\sigma _{\bar D N} = 10$ mb 
obtained in the  theoretical calculation of Ref.~\cite{Haidenbauer:2008ff}.
Besides, the averaged nuclear
density, $\bar{\rho}$, is approximated by that of 
the ${}^{11}\rm{B}$ nucleus (Eq.~\eqref{eq:rho_pn}).  By performing
the spin sums, the amplitude $\tau _{f}$ can be written as,
\begin{equation}
\label{tau}
\tau _{f} ( \bm{r} ) =
\chi _{f}^{\ast} ( \bm{r} ) \xi _{1/2,m_{s}}^{\ast}
     [ Y_{l_{\phi}}^{\ast} ( \hat{r} ) \otimes \psi _{j_{N}} ( \bm{r} )
     ]_{J M} \, \chi _{i} ( \bm{r} ) , 
\end{equation}
where it appears also the $D-$meson angular wave function
$Y_{l_{\phi}} ( \hat{r} )$, which 
 depends on the direction of the vector $\bm{r}$ ($\hat{r}$), and the
spin wave function $\xi _{1/2, m_{s}}$ of the outgoing $\bar{D}-$meson.  We assume
harmonic oscillator wave functions for the proton-hole $\psi_{j_{N}}$
wave function calculated with an
empirical value of range parameter. We stress that within this
approach, the $D$--nucleus optical potential only enters in the
Hamiltonian $H_{D}$ that appears in the Green's function.

\section{Numerical Results}
\label{sec:results}

Next, we show our numerical results for the structure and formation
spectra of the $D^{-}-$and $D^{0}-$nucleus bound states.  In
the present calculation, we focus on the ${}^{12} \rm{C} (
\bar{p}, D^{+})$ and ${}^{12} \rm{C} ( \bar{p}, \bar{D}^{0})$
reactions, and thus we consider the $[D^{-}-{}^{11}\rm{B}]$ and
$[D^{0}-{}^{11}\rm{B}]$ systems.
After discussing the properties
of the charmed meson--nucleus bound states obtained by solving  the
KGE in Sec.~\ref{sec:str}, we show the formation
spectra for these states in Sec.~\ref{sec:spec-results}. 

\subsection{Atomic and nuclear charmed meson bound states in ${}^{11}\rm{B}$}
\label{sec:str}

Binding energy $(B_{\rm E}> 0)$ and width
$(\Gamma)$ of the charmed meson--${}^{11}\rm{B}$ bound states
are related to the eigenenergy  appearing in Eq.~(\ref{KG_eq}) by $E_D = \mu -
B_{\rm E} - i \Gamma / 2$.  Since 
the $D^{-}$ or $\bar{D}^{0}$ meson optical potential
$V_{\rm opt} ( r , E_D )$ depends on the energy, we have
self-consistently solved the
KGE~\cite{GarciaRecio:2010vt, GarciaRecio:2011xt}.
\begin{table}[!t]
\begin{center}
  \caption{Binding energies $B_{\rm E}$ and widths
    $\Gamma$ of $D^-$ atomic states in ${}^{11}\rm{B}$.}
\label{tab:atom}       
\begin{tabular}{c|c|cc}
\hline\hline
State & Coulomb only & \multicolumn{2}{|c}{Coulomb $+$ optical}\\
 & $B_{\rm E}$ [keV]  & $B_{\rm E}$ [keV] & $\Gamma$ [keV] \\ \hline
$1s$ & 844.8 & 446.1 & 77.0\phantom{0} \\
$2s$ & 236.1 & 162.5 & 17.0\phantom{0} \\
$3s$ & 108.9 & \phantom{0}83.7 & 6.3 \\
$4s$ & \phantom{0}62.4  & \phantom{0}50.9 & 3.0\\
$2p$ & 264.4 & 249.7 & 9.7\\
$3p$ & 117.4 & 112.3 & 3.3\\
$4p$ & \phantom{0}66.0  & \phantom{0}63.8 & 1.5\\
$3d$ & 117.6 & 117.6 & 6.0\\
$4d$ & \phantom{0}66.1  & \phantom{0}66.1 & 3.5\\
$4f$ & \phantom{0}66.0  & \phantom{0}66.0 & 2.5\\
\hline\hline
\end{tabular}
\end{center}
\end{table}
We start by discussing the $D^{-}$ atomic levels in ${}^{11}\rm{B}$, which
are Coulomb assisted bound states.  The found levels are compiled in
Table~\ref{tab:atom}, where we show both the results obtained only with the
Coulomb potential and those obtained when the optical potential is added to
the Coulomb interaction.  We can see that the inclusion of the strong
interaction leads to smaller binding energies for both $s$ and $p$ orbital
states compared to the corresponding values obtained when only the Coulomb
interaction is considered.  This is to say, the strong interaction between the
$D^{-}-$meson and the ${}^{11}\rm{B}$ nucleus is repulsive in this case.
This is caused by the level repulsion induced by the existence of nuclear
bound $s$ and $p$ states. In addition, the imaginary part of the optical
potential has a well-known repulsive effect, also seen in
\cite{GarciaRecio:2011xt}. We emphasize that the decay widths are smaller than
the binding energies thanks to the absence of strong decay channels for the
$D^{-} N$ pair\footnote{Indeed, the small widths in the medium are due to the
  excitation of particle-hole, i.e., $\bar D \to \bar D N N^{-1}$
  \cite{GarciaRecio:2011xt}.}, which implies that such $D^{-}$ atomic states
may  be observed in dedicated experiments.
\begin{table}[!t]
\begin{center}
\caption{Binding energies $B_{\rm E}$ and widths
    $\Gamma$ of $D^-$ and $D^{0}$ nuclear states in ${}^{11}\rm{B}$. }
\label{tab:nuc}       
\begin{tabular}{c|cc|cc}
\hline
\hline
& \multicolumn{2}{|c}{$D^-$ meson}
& \multicolumn{2}{|c}{$D^0$ meson}
\\
State & $B_{\rm E}$ [MeV] & $\Gamma$ [MeV]
& $B_{\rm E}$ [MeV] & $\Gamma$ [MeV] \\
\hline
$1s$ & 21.7 & 0.5 & 6.5 & 10.8 \\
$2p$ & 14.5 & 2.4 & --- & --- \\ \hline\hline
\end{tabular}
\end{center}
\end{table}

Next, we search for charmed meson-nuclear bound states originated from the
strong interaction via the optical potential. Binding energies and decay
widths are listed in Table~\ref{tab:nuc}. We find $1s$ and $2p$ nuclear states
for $[D^{-}$-$^{11}\rm{B}]$ and a $1s$ nuclear state for the $[{D}^{0}$-$^{11}\rm{B}]$ system.  The binding energies turn out to be around ten MeV or
more.  We find narrow widths for the $D^{-}$ mesic nuclear
states\footnote{Note that it is expected that $\bD^0$ nuclear states will
  resemble those of $D^-$ due to isospin symmetry, as seen in
  Ref.~\cite{GarciaRecio:2011xt}.}, however the decay width of the $D^{0}$
mesic nuclear state is larger than its binding energy. This is due to the
existence of open strong decay modes ($\Sigma_c \pi, \Lambda_c \pi$) of the $D
N$ pair.  The natural question that arises is whether these mesic nuclear
states will appear in the spectrum of the one proton pick-up reactions.  We
address this issue in the next subsection.

\subsection{Formation spectra}
\label{sec:spec-results}

\begin{figure}[!t]
  \includegraphics[width=8.6cm]{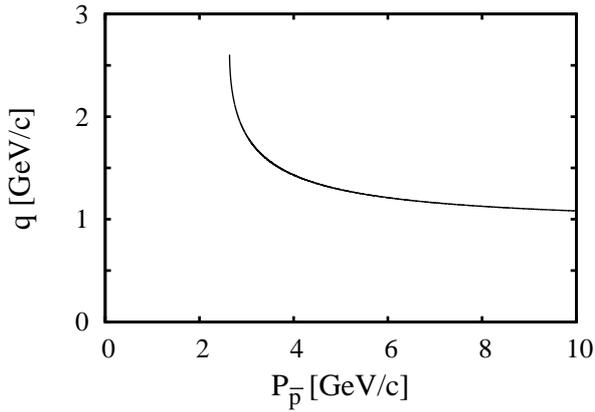}
  \caption{\label{fig:qtrans} Momentum transfer (LAB frame) as a function of the
    antiproton momentum in the ${}^{12} \rm{C} ( \bar{p} , \bar{D}
    )$ reaction. }
\end{figure}

We calculate the formation spectra of the $[D^{-}$-$^{11}\rm{B}]$ and
$[{D}^{0}$-$^{11}\rm{B}]$ systems.  We produce these states in
the ${}^{12} \rm{C} ( \bar{p} , D^{+} )$ and ${}^{12} \rm{C} (
\bar{p} , \bar{D}^{0} )$ reactions, respectively (Eqs.~(\ref{eq:rea1})
and (\ref{eq:rea2})).  We consider forward scattering for the outgoing
$D^{+}$ or $\bar{D}^{0}$ meson to maximally suppress the momentum
transferred to the mesic nuclear or atomic bound states.  Using this
kinematics, we show in Fig.~\ref{fig:qtrans} the momentum transfer in
these reactions as a function of the antiproton momentum,
$P_{\bar{p}}$, in the LAB frame.  We see  that a
large momentum transfer about $1 \gev/c$ is inevitable when working
with an antiproton beam.    

In this preliminary study we fix the LAB antiproton momentum to $8 \gev /c$,
since we expect to obtain in this region, both a large elementary cross
section~\cite{Kaidalov:1994mda} and a momentum transfer close to the smallest
possible, as seen in Fig.~\ref{fig:qtrans}.
On the other hand, we use the theoretical results of 
Ref.~\cite{Haidenbauer:2014rva} 
for the differential cross section of the elementary process
$( d \sigma / d\Omega )^{\rm LAB}$ at forward angle of the outgoing $D^{+}/
\bar{D}^{0}$ meson. Thus, at
$P_{\bar p} = 8 \gev /c$, we take 760 nb/sr (40 nb/sr) for the $p{\bar p}\to
D^+D^-$ ($p{\bar p}\to D^0{\bar D^0}$) reaction.
\begin{figure}[!t]
\includegraphics[width=8.6cm,height=6.0cm]{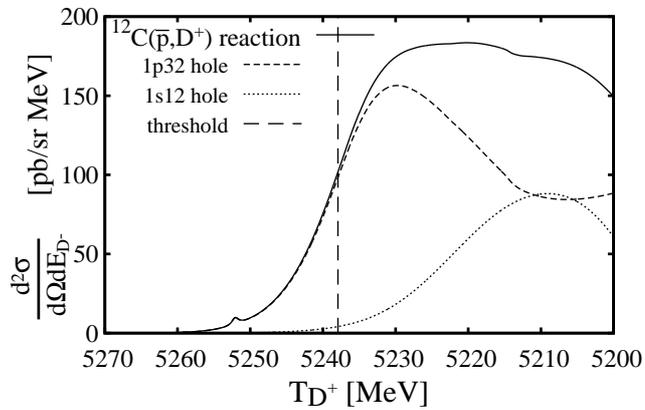}
\caption{\label{fig:Dm} Formation spectrum for the $\bar{p} + {}^{12}\rm{C}
  \to \left[ {}^{11}\rm{B}-D^- \right]+D^+$ reaction at $P_{\bar p}=8 \gev /
  c$ and $\theta^{\rm LAB}_{D^+} = 0^{\circ}$, as a function of the outgoing
  $D^{+}$ meson total energy.  The partial contributions of some shell
  configurations of the final nucleus are also shown in the figure. The
  vertical dashed line indicates the ${D^-}$ meson production threshold.}
\end{figure}

In Fig.~\ref{fig:Dm}, we show the formation spectrum of the $[D^{-}$-${}^{11}\rm{B}]$ system as a function of the outgoing $D^{+}$ meson total energy
($T_{D^{+}}$) in the LAB frame. We can appreciate a bump structure around
$T_{D^{+}} = 5250 \mev$ placed below the $D^{-}$ production threshold.  It
comes from the contribution of the $1 p_{3/2}$ hole configuration of ${}^{11}\rm{B}$, and the peak corresponds to the $2 p$ state of the $[D^{-}$-${}^{11}\rm{B}]$ nuclear state found in the previous subsection.  However, its
strength is very small compared to the quasifree contribution above the
$D^{-}$ production threshold ($T_{D^{+}} > 5238 \mev$), mainly due to the very
large momentum transfer in the reaction. The cross section for the reaction is
proportional to $|\Phi_{nlm}(q)|^2$, where $\Phi_{nlm}(q)$ is the $D^{-}$
bound wave function in momentum
space~\cite{Nieves:1990xy,Hirenzaki:1991us,Nieves:1991yf,Itahashi:1999qb}. The
cross sections are small because the $D^{-}$ bound wave function has
difficulty to accommodate such large momentum of around 1 GeV. Given the
expected typical size of the $D^{-}$ bound states (similar to $D^0$ in
Ref.~\cite{GarciaRecio:2010vt}), momentum components significantly larger than
0.4-0.5 GeV in the $D^-$ bound wave function are already expected to be
small. Thus, one gets a large suppression form factor.  The large momentum
transfer also leads to the disappearance of the $1s$ state below the $2 p$
state in the formation spectrum, since the $1s_{1/2}$ hole contribution is
negligible in the bound region\footnote{Due to the large momentum transfer,
  the convergence of the formation spectrum above the threshold becomes very
  slow and we have needed to sum up to fifteen $D^{-}$-nucleus partial
  waves.}.

%
\begin{figure}[!t]
\includegraphics[width=8.6cm,height=6.0cm]{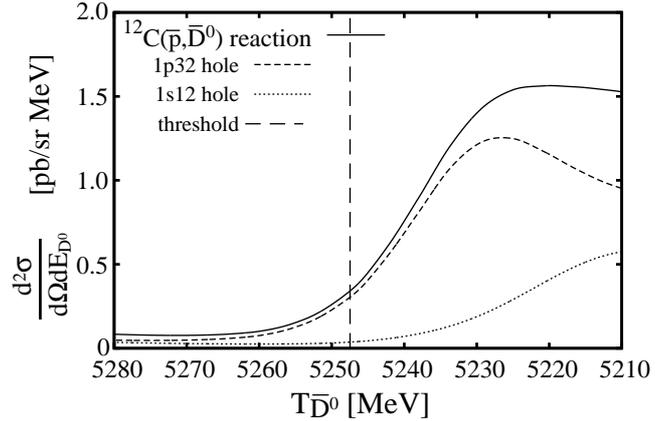}
\caption{\label{fig:D0} Same as in Fig.~\ref{fig:Dm}, but for the $\bar{p} +
  {}^{12}\rm{C} \to \left[ {}^{11}\rm{B}-D^0 \right]+\bar D^0$
  reaction. The vertical dashed line indicates now the $D^0$ meson production
  threshold.}
\end{figure}

In Fig.~\ref{fig:D0} we show now the formation spectrum of the $[D^{0}$-${}^{11}\rm{B}]$ system as a function of the outgoing $\bar{D}^{0}$ meson total
energy ($T_{\bar{D}^{0}}$) in the LAB frame.  In this case, we do not see in
the formation spectrum any signature of the $1s$ nuclear state with $B_E=6.5
\mev$ reported in Table~\ref{tab:nuc}. This is again because of the large
momentum transfer of the reaction.
\begin{figure}[!t]
\includegraphics[width=8.6cm,height=6.0cm]{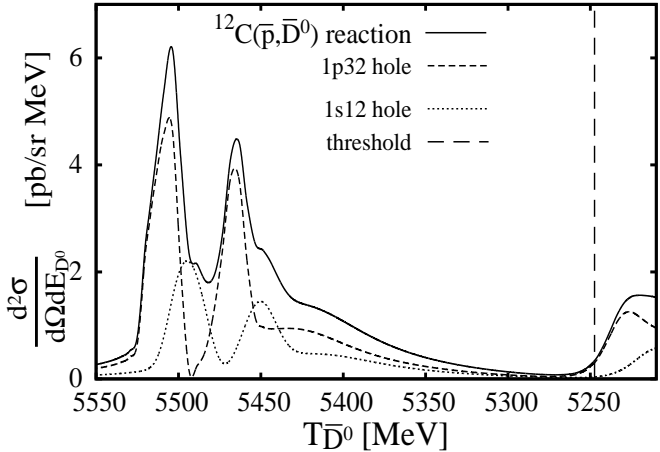}
  \caption{\label{fig:deep} As in Fig.~\ref{fig:D0}, but covering the
    region of very deep $D^0$ binding energies.}
\end{figure}
\begin{figure}[!t]
  \includegraphics[width=8.6cm]{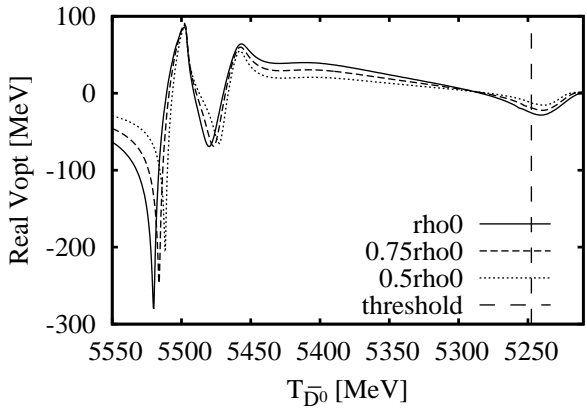}
  \caption{\label{fig:opt_deep} $D^{0}$-${}^{11}\rm{B}$ optical
    potential~\cite{Tolos:2009nn}  for several  nuclear densities in terms of the nuclear matter saturation density, $\rho_0$, as a function of 
    the $D^0$ meson energy.}
\end{figure}
However, we note that we find two peaks in formation spectrum for $\bar D^0$
energies in the region of 5.5 GeV, which would correspond to very deeply bound
$D^0$ states. These structures are shown in Fig.~\ref{fig:deep}, but we do not
obtain any eigenstates in such deep energy region.  These peaks in the
formation spectrum come from the energy dependence of the optical potential,
as can be seen in Fig.~\ref{fig:opt_deep}, and they are associated to the
dynamically-generated $\Sigma _{c} (2556)$-hole and $\Lambda _{c} (2595)-$hole
states, as discussed in \cite{Tolos:2009nn}.

As already noted, the rather large momentum transfer involved in the
$\bar{p}+p \to D + \bar{D}$ reactions tends to hinder the formation process.
Much smaller momentum transfers can be achieved with alternative reactions. 
For instance,
\begin{eqnarray}
\label{reaction}
&&\bar p + p \to D^{*-} + D^{+}, \nonumber \\
&& D^{*-} + A_Z \to \pi^0 + \left[ A_Z- D^-\right]_{\rm b}.
\end{eqnarray}
After emission of a pion the charmed meson can be slow and get trapped by the
nucleus.  More generally, in reactions of the type $\bar{p} + N \to \bar{D}^* +
D$ followed by $\bar{D}^* \to \bar{D} + \pi$ or $\bar{p} + N \to \bar{D} +
D^*$ followed by $D^* \to D + \pi$, the vector meson may be real or virtual
and the $D$ or $\bar{D}$ produced may be slow and get trapped. (Note that
nothing prevents the antiproton to annihilate with the neutrons instead of the
protons of the nucleus, thereby increasing the reaction cross section.)
Likewise, bremsstrahlung of pions produced by the antiproton as it impacts the
nucleus also changes the kinematics and could lead to new formation
mechanisms. All these alternative mechanisms in which energy and momentum is
released by emission of pions (or even photons) could help to reduce the
momentum transfer to the final charmed meson and are therefore worth studying.
From the theoretical point of view we would expect sizeable formation peaks
over a flat background.

\section{Conclusions}
\label{sec:conc}

In this work we have calculated the formation spectra of the
charmed/anti-charmed mesic nuclear states in the ${}^{12} \rm{C} ( \bar{p},
D^{+})$ and ${}^{12} \rm{C} ( \bar{p}, \bar{D}^{0})$ antiproton reactions,
aiming to provide useful information for experiments such as PANDA in the
future FAIR facility and J-PARC.  There exists also the possibility of
observing these exotic mesic nuclei in relativistic heavy-ion collisions, such
as those taking place in the future CBM experiment at FAIR.

For this purpose, we have described the (anti-)charmed meson--nucleon
scattering amplitude in dense matter by employing a unitarized
coupled-channels model based on an extended WT interaction to account for HQSS
constraints in the charm sector.  Then, with the (anti-)charmed meson--nucleon
amplitude in matter, we have constructed the self-energies and, hence, optical
potentials of the (anti-)charmed mesons in ${}^{11}\rm{B}$.  Solving the
KGE with the $D^{-}$ and $D^{0}$ optical potentials in ${}^{11}\rm{B}$, we
have found $1 s$ and $2 p$ nuclear states for the $D^{-}$ case and only the $1
s$ level for the $D^{0}$ case, in addition to the atomic states for $D^{-}$.
Of special interest is the fact that the anti-charmed $[D^{-}$-${}^{11}\rm{B}]$ nuclear states have very small decay widths from quasi-elastic
$D^{-} N\to \bar D N'$ collisions.

Next, we have calculated the charmed and anti-charmed mesic nuclear formation
spectra in the ${}^{12}C ( \bar{p}, D^{+})$ and ${}^{12}C ( \bar{p},
\bar{D}^{0})$ reactions by employing the Green's function method.  The
momentum of the antiproton beam has been fixed to $8 \gev / c$, and the final
mesons $D^{+}/\bar{D}^{0}$ are taken in the forward direction to suppress as
much as possible the momentum transferred to the mesic nuclear bound states.
We have found, on the one hand, that the $2 p$ $D^{-}$ nuclear state may show
up in the formation spectrum as a small peak.  However, its peak strength is
very small compared to the quasifree contribution above the $D^{-}$ production
threshold, mainly due to the very large momentum transfer in the reaction.
The large momentum transfer also leads to the disappearance of the $1 s$ state
below the $2 p$ state in the formation spectrum for $D^{-}$.  On the other
hand, for the $D^{0}$ meson, the nuclear bound state does not lead to any
visible signatures in the spectrum, although at deep energy regions large
peaks are present. These structures correspond to the dynamically-generated
$\Sigma _{c} (2556)$-hole and $\Lambda _{c} (2595)$-hole excitations, and
their experimental observation might shed light into the dynamics of these
resonances inside of a nuclear environment.

Finally we note that in the $( \bar{p} , D^{+} )$ and $( \bar{p} , \bar{D}^{0}
)$ reactions, the momentum transfer is inevitably large.  Therefore, in order
to have visible strengths for the $1 s$ or $2 p$ nuclear states, we should
consider different production reactions with small momentum transfer.  One
possibility is to examine the $( \bar{p} , D + N )$ and $( \bar{p} , D + 2 N
)$ reactions, with a much smaller or even zero momentum transfer, although the
formation cross sections could be suppressed as well because of the complexity
of the reaction mechanisms. Other competing formation mechanisms could involve the emission of pions by real or virtual intermediate $D^*$ or $\bar{D}^*$ with subsequent trapping of the slow pseudoscalar charmed meson by the final nucleus.

\appendix
\section{Neutron and proton densities}
\label{appendix}

For the evaluation of the nuclear density $\rho (r)$ of a given nucleus, one
has to obtain the neutron and proton densities.  Namely for the
${}^{11}\rm{B}$ nucleus, on the one hand, the charge distribution $\rho_{\rm
  ch}$ is given by a modified harmonic oscillator (MHO) distribution
\begin{equation}
  \rho_{\rm{ch}} ( r ) = \displaystyle \rho_{0}
  \left [ 1 + a \left ( \frac{r}{R} \right ) ^{2} \right ]
  \exp \left [ - \left ( \frac{r}{R} \right ) ^{2} \right ] ,
  \label{rho_ch}
\end{equation}
The neutron matter distribution is taken to be identical to $\rho_{\rm ch}$.
On the other hand, the densities of proton ($\rho_{p}$) and neutron
($\rho_{n}$) turn out to have also a MHO shape~\cite{Nieves:1993ev}, but with
modified parameters to account for the proton and neutron finite sizes,

\begin{equation}
  \rho_{p, n} ( r ) = \displaystyle \rho_{0}^{\prime}
  \left [ 1 + a^{\prime} \left ( \frac{r}{R^{\prime}} \right ) ^{2} \right ]
  \exp \left [ - \left ( \frac{r}{R^{\prime}} \right ) ^{2} \right ] ,
\end{equation}
with the parameters $a^{\prime}$ and $R^{\prime}$ given by
\begin{eqnarray}
  a^{\prime} &=& \frac{2 x}{2 - 3 x} ,
  \quad
  x = \frac{a R^{2}}{1 + 3 a / 2} \frac{1}{R^{\prime 2}},\\  
  R^{\prime} &=& \sqrt{ R^{2} - \frac{2}{3} \langle r_{p, n} \rangle ^{2} } ,
\end{eqnarray}
with the mean radius of the proton or neutron $\langle r_{p, n}^2
\rangle=0.69$ fm$^2$.  Then the nuclear density $\rho$ is the sum of the
proton and neutron densities:
\begin{eqnarray}
  \rho ( r ) = \rho _{p} ( r ) + \rho _{n} ( r ) .
  \label{eq:rho_pn}
\end{eqnarray}
We use $R=1.69 \fm$ and $a=0.811$, and the normalization factors $\rho _{0}$ and $\rho
_{0}^{\prime}$ are determined so that $\int d^{3} r \rho _{\rm ch} ( r
) = \int d^{3} r \rho _{p} ( r ) = 5$, and $\int d^{3} r \rho _{n} ( r
) = 6$.

\noindent
\rm{\bf{Acknowledgments}}
We acknowledge the support by Open Partnership Joint Projects of JSPS
Bilateral Joint Research Projects. This research has been supported by the
Spanish Ministerio de Econom\'\i a y Competitividad and European FEDER funds
under the contracts FIS2011-28853-C02-02, FPA2013-43425-P,
FIS2014-51948-C2-1-P, FIS2014-59386-P, FIS2014-57026-REDT and
SEV-2014-0398. Also, this work has been financed by Generalitat Valenciana
under contract PROMETEOII/2014/0068, by Junta de Andaluc\'{\i}a (grant
FQM225), and by the EU HadronPhysics3 project, grant agreement no. 283286.
LT acknowledges support from the Ram\'on y Cajal Research Programme from Ministerio de Econom\'ia y Competitividad.




\begin{thebibliography}{99}



\bibitem[()]{Ericson:1966fm}
  M.~Ericson, T.~E.~O.~Ericson,
  Annals Phys.\  {\bf 36} (1966) 323.

\bibitem[()]{Friedman:2007zza}
  E.~Friedman, A.~Gal,
  Phys.\ Rept.\  {\bf 452} (2007) 89.

\bibitem[()]{Nieves:1993ev} 
  J.~Nieves, E.~Oset and C.~Garcia-Recio,
  Nucl.\ Phys.\ A {\bf 554} (1993) 509.

\bibitem[()]{GarciaRecio:1991wk}
  C.~Garcia-Recio, J.~Nieves, E.~Oset,
  Nucl.\ Phys.\  A {\bf 547} (1992) 473.


\bibitem[()]{Hirenzaki:2008zz}
  S.~Hirenzaki,
  Mod.\ Phys.\ Lett.\  A {\bf23} (2008) 2497.
  
\bibitem[()]{Gilg:1999qa}
  H.~Gilg, A.~Gillitzer, M.~Knulle, M.~Munch, W.~Schott, P.~Kienle, K.~Itahashi, K.~Oyama {\it et al.},
  Phys.\ Rev.\  C {\bf 62} (2000) 025201.

\bibitem[()]{Wycech:2007jb}
  S.~Wycech, F.~J.~Hartmann, J.~Jastrzebski, B.~Klos, A.~Trzcinska, T.~von Egidy,
  Phys.\ Rev.\  C {\bf 76} (2007) 034316.

\bibitem[()]{Klos:2007is}
  B.~Klos, A.~Trzcinska, J.~Jastrzebski, T.~Czosnyka, M.~Kisielinski, P.~Lubinski, P.~Napiorkowski, L.~Pienkowski {\it et al.},
  Phys.\ Rev.\  C {\bf 76} (2007) 014311.

\bibitem[()]{Trzcinska:2001sy}
  A.~Trzcinska, J.~Jastrzebski, P.~Lubinski, F.~J.~Hartmann, R.~Schmidt, T.~von Egidy, B.~Klos,
  Phys.\ Rev.\ Lett.\  {\bf 87} (2001) 082501.

  

\bibitem[()]{fair} http://www.fair-center.eu/public/experiment-program.html

\bibitem[()]{jparc}http://j-parc.jp/researcher/Hadron/en/index.html



\bibitem[()]{Tsushima:1998ru} 
  K.~Tsushima, D.~H.~Lu, A.~W.~Thomas, K.~Saito and R.~H.~Landau,
  Phys.\ Rev.\ C {\bf 59} (1999) 2824.

\bibitem[()]{Guichon:1987jp}
  P.~A.~M.~Guichon,
  Phys.\ Lett.\  B {\bf 200} (1988) 235.

\bibitem[()]{Yasui:2012rw} 
  S.~Yasui and K.~Sudoh,
  Phys.\ Rev.\ C {\bf 87} (2013) 015202.

\bibitem[()]{Yasui:2009bz} 
  S.~Yasui and K.~Sudoh,
  Phys.\ Rev.\ D {\bf 80} (2009) 034008.
  
\bibitem[()]{Yamaguchi:2013hsa} 
  Y.~Yamaguchi, S.~Yasui and A.~Hosaka,
  Nucl.\ Phys.\ A {\bf 927} (2014) 110.

\bibitem[()]{Bayar:2012dd}
  M.~Bayar, C.~W.~Xiao, T.~Hyodo, A.~Dote, M.~Oka and E.~Oset,
  {Phys.\ Rev.\ C} {\bf 86} (2012) 044004.
  


  
\bibitem[()]{Tolos:2004yg}
  L.~Tolos, J.~Schaffner-Bielich and A.~Mishra,
 { Phys.\ Rev.\  C} {\bf 70} (2004) 025203.

\bibitem[()]{Tolos:2005ft} 
  L.~Tolos, J.~Schaffner-Bielich and H.~Stoecker,
 { Phys.\ Lett.\ B} {\bf 635}  (2006) 85. 

\bibitem[()]{Lutz:2003jw}
  M.~F.~M.~Lutz and E.~E.~Kolomeitsev,
 { Nucl.\ Phys.\  A}  {\bf 730} (2004) 110.  

\bibitem[()]{Lutz:2005ip} 
  M.~F.~M.~Lutz and E.~E.~Kolomeitsev,
  { Nucl.\ Phys.\ A}  {\bf 755} (2005) 29.  

\bibitem[()]{Hofmann:2005sw} 
  J.~Hofmann and M.~F.~M.~Lutz,
  { Nucl.\ Phys.\ A} {\bf 763} (2005) 90. 

\bibitem[()]{Hofmann:2006qx}
  J.~Hofmann and M.~F.~M.~Lutz,
 { Nucl.\ Phys.\  A} {\bf 776} (2006) 17. 

\bibitem[()]{Lutz:2005vx} 
  M.~F.~M.~Lutz and C.~L.~Korpa,
  { Phys.\ Lett.\ B} {\bf 633} (2006) 43.  

\bibitem[()]{Mizutani:2006vq} 
  T.~Mizutani and A.~Ramos,
  { Phys.\ Rev.\ C} {\bf 74} (2006) 065201.   

\bibitem[()]{Tolos:2007vh}
  L.~Tolos, A.~Ramos and T.~Mizutani,
  { Phys.\ Rev.\  C} {\bf 77} (2008) 015207. 

\bibitem[()]{JimenezTejero:2009vq} 
  C.~E.~Jimenez-Tejero, A.~Ramos and I.~Vidana,
  { Phys.\ Rev.\ C} {\bf 80} (2009) 055206.  
  
\bibitem[()]{JimenezTejero:2011fc} 
  C.~E.~Jimenez-Tejero, A.~Ramos, L.~Tolos and I.~Vidana,
  Phys.\ Rev.\ C {\bf 84} (2011) 015208.


\bibitem[()]{Haidenbauer:2007jq}
  J.~Haidenbauer, G.~Krein, U.~G.~Meissner and A.~Sibirtsev,
  { Eur.\ Phys.\ J.\  A} {\bf 33}  (2007) 107. 

\bibitem[()]{Haidenbauer:2008ff}
  J.~Haidenbauer, G.~Krein, U.~G.~Meissner and A.~Sibirtsev,
  { Eur.\ Phys.\ J.\  A} {\bf 37} (2008) 55.

\bibitem[()]{Haidenbauer:2010ch}
  J.~Haidenbauer, G.~Krein, U.~G.~Meissner and L.~Tolos,
  { Eur. Phys. J A} {\bf 47} (2011) 18. 

\bibitem[()]{Wu:2010jy}
  J.~-J.~Wu, R.~Molina, E.~Oset and B.~S.~Zou,
 {  Phys.\ Rev.\ Lett.}  {\bf 105} (2010) 232001. 
 
\bibitem[()]{Wu:2010vk} 
  J.~-J.~Wu, R.~Molina, E.~Oset and B.~S.~Zou,
  { Phys.\ Rev.\ C} {\bf 84}  (2011) 015202. 
 
\bibitem[()]{Wu:2012md} 
  J.~-J.~Wu, T.~-S.~H.~Lee and B.~S.~Zou,
 {  Phys.\ Rev.\ C}  {\bf 85}  (2012) 044002. 

\bibitem[()]{Oset:2012ap} 
  E.~Oset, A.~Ramos, E.~J.~Garzon, R.~Molina, L.~Tolos, C.~W.~Xiao, J.~J.~Wu and B.~S.~Zou,
 {  Int.\ J.\ Mod.\ Phys.\ E}  {\bf 21}  (2012) 1230011. 
  
  


\bibitem[()]{IW89} N. Isgur and M.B. Wise, Phys. Lett. B {\bf 232} (1989) 113.

\bibitem[()]{Ne94}  M. Neubert, Phys. Rep. {\bf 245} (1994) 259.

\bibitem[()]{MW00} A.V. Manohar and M.B. Wise, {\it Heavy Quark Physics},
  Cambridge Monographs on Particle Physics, Nuclear Physics and
  Cosmology, vol. 10

 

\bibitem[()]{GarciaRecio:2008dp} 
  C.~Garcia-Recio, V.~K.~Magas, T.~Mizutani, J.~Nieves, A.~Ramos, L.~L.~Salcedo and L.~Tolos,
  Phys.\ Rev.\ D {\bf 79} (2009) 054004.
  
\bibitem[()]{Gamermann:2010zz} 
  D.~Gamermann, C.~Garcia-Recio, J.~Nieves, L.~L.~Salcedo and L.~Tolos,
  Phys.\ Rev.\ D {\bf 81} (2010) 094016.

\bibitem[()]{Romanets:2012hm} 
  O.~Romanets, L.~Tolos, C.~Garcia-Recio, J.~Nieves, L.~L.~Salcedo and R.~G.~E.~Timmermans,
  Phys.\ Rev.\ D {\bf 85} (2012) 114032.
  
\bibitem[()]{GarciaRecio:2012db} 
  C.~Garcia-Recio, J.~Nieves, O.~Romanets, L.~L.~Salcedo and L.~Tolos,
  Phys.\ Rev.\ D {\bf 87} (2013) 034032.
  
\bibitem[()]{Garcia-Recio:2013gaa} 
  C.~Garcia-Recio, J.~Nieves, O.~Romanets, L.~L.~Salcedo and L.~Tolos,
  Phys.\ Rev.\ D {\bf 87} (2013) 074034.
  
\bibitem[()]{Tolos:2013gta} 
  L.~Tolos,
  Int.\ J.\ Mod.\ Phys.\ E {\bf 22} (2013) 1330027.

\bibitem[()]{Xiao:2013yca} 
  C.~W.~Xiao, J.~Nieves and E.~Oset,
  Phys.\ Rev.\ D {\bf 88} (2013) 056012.

\bibitem[()]{Ozpineci:2013zas} 
  A.~Ozpineci, C.~W.~Xiao and E.~Oset,
  Phys.\ Rev.\ D {\bf 88} (2013) 034018.


  
\bibitem[()]{Tolos:2009nn} 
  L.~Tolos, C.~Garcia-Recio and J.~Nieves,
  Phys.\ Rev.\ C {\bf 80} (2009) 065202.


\bibitem[()]{GarciaRecio:2010vt} 
  C.~Garcia-Recio, J.~Nieves and L.~Tolos,
  Phys.\ Lett.\ B {\bf 690} (2010) 369.

\bibitem[()]{GarciaRecio:2011xt} 
  C.~Garcia-Recio, J.~Nieves, L.~L.~Salcedo and L.~Tolos,
  Phys.\ Rev.\ C {\bf 85} (2012) 025203.
  



\bibitem[()]{Gamermann:2011mq} 
  D.~Gamermann, C.~Garcia-Recio, J.~Nieves and L.~L.~Salcedo,
  Phys.\ Rev.\ D {\bf 84} (2011) 056017.


\bibitem[()]{Morimatsu:1985pf} 
  O.~Morimatsu and K.~Yazaki,
  Nucl.\ Phys.\ A {\bf 435} (1985) 727.

\bibitem[()]{Hayano:1998sy} 
  R.~S.~Hayano, S.~Hirenzaki and A.~Gillitzer,
  Eur.\ Phys.\ J.\ A {\bf 6} (1999) 99.
  
\bibitem[()]{Klingl:1998zj} 
  F.~Klingl, T.~Waas and W.~Weise,
  Nucl.\ Phys.\ A {\bf 650} (1999) 299.

\bibitem[()]{Jido:2002yb} 
  D.~Jido, H.~Nagahiro and S.~Hirenzaki,
  Phys.\ Rev.\ C {\bf 66} (2002) 045202.

\bibitem[()]{Nagahiro:2003iv} 
  H.~Nagahiro, D.~Jido and S.~Hirenzaki,
  Phys.\ Rev.\ C {\bf 68} (2003) 035205.

\bibitem[()]{Nagahiro:2005gf} 
  H.~Nagahiro, D.~Jido and S.~Hirenzaki,
  Nucl.\ Phys.\ A {\bf 761} (2005) 92.

\bibitem[()]{Haidenbauer:2014rva} 
  J.~Haidenbauer and G.~Krein,
  Phys.\ Rev.\ D {\bf 89} (2014) 114003.



\bibitem[()]{Kaidalov:1994mda} 
  A.~B.~Kaidalov and P.~E.~Volkovitsky,
  Z.\ Phys.\ C {\bf 63} (1994) 517.
  
\bibitem[()]{Nieves:1990xy} 
  J.~Nieves and E.~Oset,
  Nucl.\ Phys.\ A {\bf 518} (1990) 617.

\bibitem[()]{Hirenzaki:1991us} 
  S.~Hirenzaki, H.~Toki and T.~Yamazaki,
  Phys.\ Rev.\ C {\bf 44} (1991) 2472.

\bibitem[()]{Nieves:1991yf} 
  J.~Nieves and E.~Oset,
  Phys.\ Lett.\ B {\bf 282} (1992) 24.
\bibitem[()]{Itahashi:1999qb}
  K.~Itahashi {\it et al.},
  Phys.\ Rev.\  C {\bf 62} (2000) 025202.



\end{thebibliography}


\end{document}